\documentclass[aps,pre,reprint,superscriptaddress]{revtex4-1}

\usepackage{graphicx}
\usepackage{amssymb}
\usepackage{amsmath}
\usepackage{bm}%
\usepackage[colorlinks=true,linkcolor=blue]{hyperref}%
\usepackage{subeqnarray}
\usepackage{gensymb}

\begin{document}


\title{Imbibition in plant seeds}


\author{Jean-Fran\c cois Louf$^\dagger$}
\affiliation{Department of Physics, Technical University of Denmark, DK-2800 Kgs. Lyngby, Denmark}
\author{Yi Zheng$^\dagger$}
\affiliation{Department of Physics, Technical University of Denmark, DK-2800 Kgs. Lyngby, Denmark}
\author{Aradhana Kumar}
\affiliation{Department of Physics, Technical University of Denmark, DK-2800 Kgs. Lyngby, Denmark}
\author{Tomas Bohr}
\affiliation{Department of Physics, Technical University of Denmark, DK-2800 Kgs. Lyngby, Denmark}
\author{Carsten Gundlach}
\affiliation{Department of Physics, Technical University of Denmark, DK-2800 Kgs. Lyngby, Denmark}
\author{Jesper Harholt}
\affiliation{Carlsberg Research Laboratory, J.C. Jacobsens Gade 4, DK-1799, Copenhagen V, Denmark}
\author{Henning Friis Poulsen}
\affiliation{Department of Physics, Technical University of Denmark, DK-2800 Kgs. Lyngby, Denmark}
\author{Kaare H. Jensen}
\affiliation{Department of Physics, Technical University of Denmark, DK-2800 Kgs. Lyngby, Denmark}
\email[]{khjensen@fysik.dtu.dk}



\date{\today}

\begin{abstract}
We describe imbibition in real and artificial plant seeds, using a combination of experiments and theory. In both systems, our experiments demonstrate that liquid permeates the substrate at a rate which decreases gradually over time. Tomographic imaging of soy seeds is used to confirmed this by observation of the permeating liquid using an iodine stain. To rationalize the experimental data, we propose a model based on capillary action which predicts the temporal evolution of the radius of the wet front and the seed mass. The depth of the wetting front initially evolves as $t^{1/2}$ in accord with the Lucas-Washburn law. At later times, when the sphere is almost completely filled, the front radius scales as $(1-t/t_{\text{max}})^{1/2}$ where $t_{\text{max}}$ is the time required to complete imbibition.
The data obtained on both natural and artificial seeds collapse onto a single curve that agrees well with our model, suggesting that  capillary phenomena contibute to moisture uptake in soy seeds.

($\dagger=$ equal contributors.)
\end{abstract}


\maketitle
\section{Introduction}
Imbibition is the spontaneous uptake of liquids by dry porous materials. It is a process which plays a key role in numerous industrial processes, for instance, in painting, printing and oil recovery \cite{Xiao2012}. 
Imbibition is also a critical stage in germination of plants seeds: it is essential for enzyme activation, breakdown of starch into sugars and transport of nutrients to the developing embryo \cite{bewley1997seed,linkies2010evolution,1461446929}.
Imbibition in plant seeds has been studied under a range of conditions, e.g. drought or salinity tolerance, yet the basic physical mechanisms that influence the rate of uptake in many seeds are not clear \cite{1461446929,swanson1985seed,nakayama2008water}.

Water is a basic requirement for germination of plant seeds.  In their resting state, plant seeds are low in moisture ($5-15$\%) and almost metabolically inactive. A remarkable property of seeds is that they are able to survive in this state, often for many years. Most seeds have a critical moisture content for germination to occur. For example, this value in corn is approximately $30\%$ while for wheat it is $50\%$ \cite{sims1959germination,ashraf1978wheat}. Once that critical seed moisture content is attained, germination starts and cannot subsequently be reversed. If the internal moisture content later decreases below the critical value, most seeds will decay in the soil. Thus, precise temporal and spatial control of the imbibition process is essential: If the process is too fast, the seed would risk initiating germination in a dry environment. If, by contrast, the uptake of water occurs slowly, the seed would risk falling behind in the competition with other plants for light and soil nutrients \cite{1461446929}. 

The duration of imbibition depends on certain inherent properties of the seed, e.g., hydratable substrate content, seed coat permeability, seed size and on the prevailing conditions during hydration: temperature, initial moisture content, water and oxygen availability. Moreover, different parts of a seed may pass through these phases at different rates; e.g., an embryo or tissue located near the surface of a large seed may swell even before its associated bulky storage tissue has become fully imbibed \cite{1461446929,montanuci2013kinetic,rathjen2009water,ha2018poro}. The structure of the porous material also influences the rate of moisture uptake. Multi-scale pores are common in biological materials \cite{gibson2012hierarchical}; either due to variations in the particle/cell size or due to the presence of cracks. It is thus unclear if the imbibition process in plant seeds is homogenous, or if spatial variations could lead to heterogeneous flow patterns \cite{kim2017capillary,debacker2014imbibition,hong2009mr,gruwel2001magnetic}. 

In this paper, we make a first attempt at separating physical and biological processes in the hydration process. Our approach is to combine imbibition experiments on real and artificial seeds with theory to elucidate the physical processes that control water uptake in plant seeds.




\section{Experiments}
\begin{figure*}
\centering
\includegraphics[width=16cm]{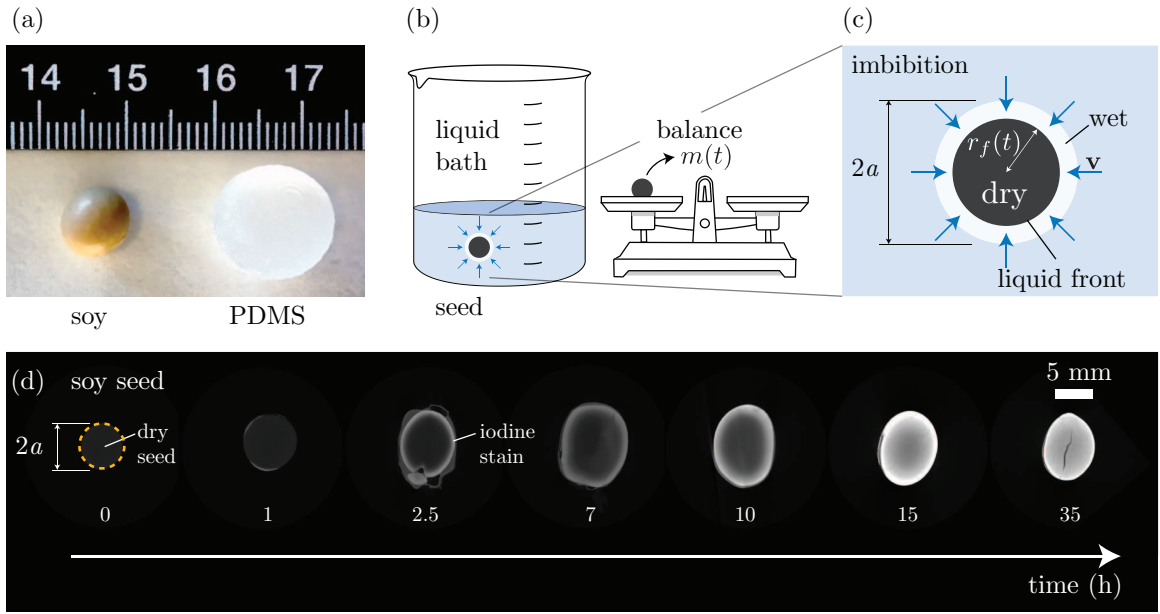}
\caption{Imbibition in plants seeds. (a) Photograph of soy and biomimetic  seeds. (b) Schematic of the experimental setup, and (c) zoom-in on the imbibition into the seed  of radius $a$. The blue arrows indicate the liquid flow speed $\mathbf v$ into the porous seed, thus gradually reducing the dry front position $r_f$. (d) X-ray tomography images of the imbibition of water into a soy seed. The change in gray-scale intensity indicates the binding of an iodine stain to the starch in the seed. The approximate rotational symmetry of the scans indicate that the imbibition process in soy seeds is homogenous. See additional details of the experimental methods in the text. \label{Fig:Intro}}
\end{figure*}

Imbibition experiments were conducted on soy (\emph{Glycine max}) and artificial seeds (Fig.~\ref{Fig:Intro}(a)). Each seed was submerged in a liquid bath, and the mass $m(t)$ was measured at regular time intervals (Fig.~\ref{Fig:Intro}(b,c)). Finally, X-ray computed tomography was used to directly visualize the imbibition process (Fig.~\ref{Fig:Intro}(d)).

\subsection{Soy seeds}
Soy seeds of diameter $d=6-8$ mm  were stored at temperatures $20,\;50$ and $80^\circ$C for 24 hours. 
For every temperature, each of the $N=7-12$ seeds were placed in separate beakers filled with tap water. The seeds were periodically weighed one-by-one using a balance (Quintix124-1S, Sartorius Lab Instruments GmbH $\&$ co, Goettingen, Germany) to quantify the change in seed mass $m(t)$ over time $t$, starting at $m(0)=m_0$. The seeds were removed from the water bath, blotted twice with a dry paper towel, and then immediately transferred to the scale. Measurements were taken regularly until the seed mass saturated; first at short, and then longer, time intervals. 
Representative experimental graphs are shown in Fig.~\ref{Fig:RawData}(a). Starting at $m_0=1.5-2.1$ g, the seed mass gradually increases over time, before reaching a plateau at $m_{\text{max}}\simeq 0.4$ g after approximately $10$ hours. 

\subsection{Artificial seeds}
An artificial seed of diameter $d=12$ mm was produced by pouring a PDMS polymer solution (Sylgard 184, Dow Europe, Germany) into a negative 3D-printed mold. The solution contained 10$\%$ (wt) cross-linker and was degassed for 60 min. The mold was fabricated from polylactic acid (PLA) on a Ultimaker 3 printer (Ultimaker B.V., The Netherlands).
The PDMS was cured in the mold for 2h at 100$\degree$C, and the mass of the seed was $m_0=0.9$ g.  The artifical seed was subsequently placed in a beaker filled with the solvent Diisopropylamine. The seed was weighed using the balance to quantify the change in seed mass $m(t)$ over time $t$, starting at $m(0)=m_0$. The seed was removed from the solvent bath, blotted twice with a dry paper towel, and then immediately transferred to the scale. Measurements were taken every hour for the first ten hours, then regularly for the next 60 hours.  A typical experiment is illustrated in Fig.~\ref{Fig:RawData}(b). Starting at $m_0$, the seed mass gradually increases over time, before reaching a plateau at $m_{\text{max}}\simeq 3.5$ g after approximately $50$ hours. The solvent-PDMS contact angle $\theta \simeq 150^\circ$ was determined following the procedure in \cite{gart2015}.


\subsection{X-ray imaging}
To elucidate the spatial dynamics of the imbibition process, $7$ different soy seeds were imaged using a ZEISS Xradia 410 Versa (Carl Zeiss X-ray Microscopy, Inc., Germany) X-ray computed tomography system. Initially, the seeds were placed in beakers filled Lugol's aqueous iodine solution. A seed was removed after $t=0,\,1,\, 2.5,\, 7,\, 10,\,15$ and $35$ hours and placed in the scanner. Imaging took $\sim 1$ hour for each seed and followed the protocol outlined in \cite{ludwig2009new}. The CT data was reconstructed using the Reconstruction in the Zeiss ?scout and scan software package? (Version 11, Carl Zeiss X-ray Microscopy, Inc., Germany), which is based on the FDK algorithm \cite{feldkamp1984practical}, using a cone beam filtered back projection approach. The global parameters were set to be the same for all data such that a given material in different scans will have identical grey values. 

\begin{figure}
\centering
  \includegraphics[width=8cm]{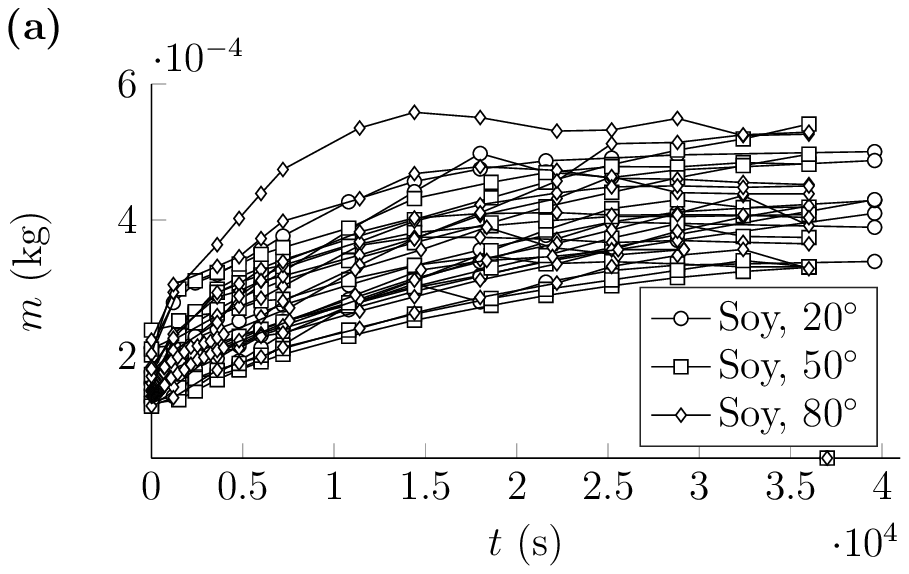}\\
    \includegraphics[width=8cm]{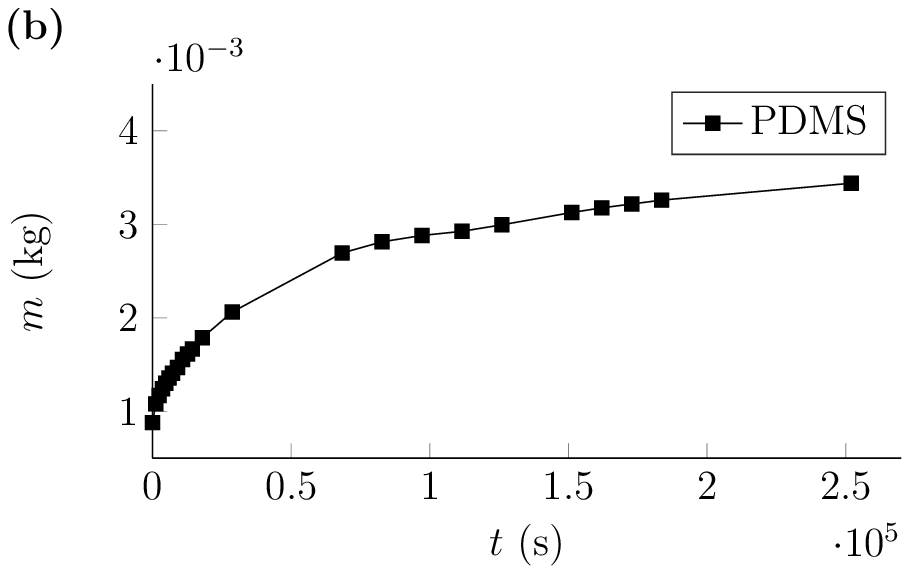}\\ 
\caption{Imbibition kinetics in real and artificial seeds. (a) Plot of the mass $m$ of soy seeds as a function of time $t$. Prior to the experiments, the seeds were stored at temperatures $20,\;50$ and $80^\circ$C for 24 hours (see legend). 
(b) Plot of the mass $m$ of a biomimetic PDMS seed as a function of time $t$. Both seeds exhibit similar behavior: the mass increases gradually until it reaches a plateau corresponding to a fully wetted sphere.\label{Fig:RawData}}
\end{figure}

The iodine solution provided sufficient contrast to qualitatively follow the propagation of the liquid front inside the seed (Fig. \ref{Fig:Intro}(d)). After a delay during which the seed coat was penetrated, liquid quickly entered the porous material. The rate of water uptake gradually reduced, in accord with Fig. \ref{Fig:RawData}(a). The approximate rotational symmetry of the scans indicate that the imbibition process in soy seeds is homogenous, suggestive of a single pore scale porous material.  Note, however, that because the binding of iodine -- and hence the observed intensity -- is a gradual process, the motion of liquid front cannot directly be imaged in real time. We thus expect a temporal delay between the arrival of the liquid front and accumulation of sufficient iodine to appear on the CT scan.



\section{Theory of imbibition in plant seeds}
To rationalize the observed imbibition dynamics (Figs. \ref{Fig:Intro}(c) and \ref{Fig:RawData}), we proceed to consider the intrusion of liquid into a dry homogenous porous material. 
Imbibition follows the ``Lucas-Washburn" law: the liquid is pulled in by the pressure gradient created by the difference between the pressure of the source (ambient pressure) and that of the front - the capillary pressure \cite{lucas1918rate,washburn1921dynamics,Xiao2012,Capillarity}. The resistance to this pressure-driven flow comes from the narrow pores in the material and is described by Darcy's law giving a linear relation between water velocity $\mathbf v$ and the local pressure gradient $\mathbf \nabla p$, typically written as
\begin{equation}
\label{Darcy}
{\bf v}= -\frac{k}{\mu} \nabla p,
\end{equation}
where $\mu$ is the viscosity of the fluid and $k\sim r_p^2$ is the ``permeability", proportional to the square of the mean radius $r_p$ of the pores.

The capillary (or Laplace) pressure is also determined by the pore-sizes, together with the surface tension $\gamma$ and the contact angle $\theta$:
\begin{equation}
\label{capil}
p_c = -\frac{2 \gamma \cos \theta}{r_p} .
\end{equation}
Imbibition in a spherical geometry has previously been studied in the context of polymer penetration into silica agglomerates \cite{bohin1994penetration,bohin1995determination}, and the process was recently reviewed by \citet{Xiao2012}. As long as there is no tissue swelling, the water imbibition takes place by filling up the pores and thus the water flow in the wet volume must be divergence free: $\mathbf \nabla \cdot {\bf v}=0$. Together with Darcy's law (\ref{Darcy}) with constant permeability $k$ and viscosity $\mu$ this implies that the pressure satisfies Laplace's equation, i.e., 
\begin{equation}
\label{Laplace}
 \nabla^2 p = 0.
\end{equation}
Assuming a spherical geometry, where ${\bf v}= v(r) \bf e_r$ is purely radial and therefore $p=p(r)$ then gives
\begin{equation}
\label{p1}
p=\frac{A}{r}+ B.
\end{equation}
Let us assume that the seed has the radius $a$ and the liquid at $r=a$ is at ambient pressure, which we shall denote $p(a)=0$. At a given time $t$ the front has reached the position $r_f(t)$, where the pressure is given by Eq.~\eqref{capil}. Thus
\begin{equation}
\label{p2}
p_c=A\left(\frac{1}{r_f} -\frac{1}{a} \right).
\end{equation}
The constant $A$ is found from Darcy's law giving
\begin{equation}
\label{p3}
v(r) = -\frac{k}{\mu}[-p'(r)]=-\frac{kA}{\mu r_f^2},
\end{equation}
where prime denotes differentiation with respect to $r$ and the speed is radially inward from the surface. The flow rate $q$ at the front (and at any radial position in the wet volume) is found by integrating the velocity over the surface of the sphere $q(t) = 4 \pi r_f^2 (-v(r))$
\begin{equation}
\label{A}
A=\frac{\mu}{4 \pi k }q
\end{equation}
giving us, from Eqns. \eqref{capil} and \eqref{p2} the relation
\begin{equation}
p_c=\frac{\mu}{4 \pi k }q\left(\frac{1}{r_f} -\frac{1}{a} \right)= -\frac{\mu}{ k }r_f^2(t) r_f'(t)\left(\frac{1}{r_f(t)} -\frac{1}{a} \right),\label{eq:pc2}
\end{equation}
where we have used that the front velocity $r_f$ is identical to the fluid velocity at the front:
\begin{equation}
\label{vf}
v(r_f) = r_f'(t).
\end{equation}
Equation \eqref{eq:pc2} can be integrated as
\begin{equation}
\label{rf}
-\frac{1}{3a}\left(a^3 - r_f^3 \right) + \frac{1}{2}\left(a^2 - r_f^2 \right) = \frac{k p_c}{\mu}t.
\end{equation}
One interesting consequence of Eq. (\ref{rf}) is that the time $t_{max}$ to complete imbibition (where $r_f=0$) is
\begin{equation}
\label{tmax}
t_{max}=\frac16 \frac{\mu }{k p_c} a^2,
\end{equation}
which is proportional to the square of the radius $a$. We also note that $t_{\text{max}}$ is inversely proportional to the pore size $r_p$, since the product $k p_c$ scales as $k p_c \sim r_p^2 r_p^{-1}=r_p$. 

\begin{figure}
\centering
\includegraphics[width=8cm]{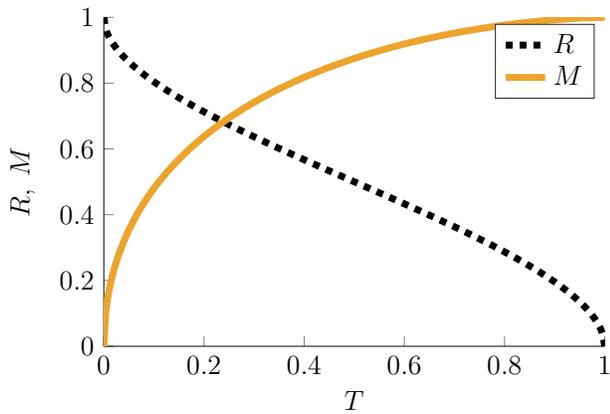}
\caption{Theoretical values of the normalized radius $R$ (Eq. \eqref{eq:RT}, dashed line) and mass $M$ (Eq. \eqref{eq:M}, solid line) plotted as as a function of the normalized time $T=t/t_{\text{max}}$.\label{Fig:theory_radius_mass}}
\end{figure}

Introducing the non-dimensional front position $R=r_f/a$ and time $T= t/t_{\text{max}}$ leads to
\begin{equation}
1-3R^2+2R^3=T,
\end{equation}
and thus there is a universal growth curve for $R(T)$ when the time is also scaled by the only natural scale. The solution to this equation is real when $0<T<1$ and is given by \cite{bohin1994penetration}
\begin{equation}
    R(T) = \frac{1}{2}-\cos \left[\frac{1}{3} \left(\cos ^{-1}[1-2 T]+4 \pi \right)\right].\label{eq:RT}
\end{equation}
Figure \ref{Fig:theory_radius_mass} shows the normalized radius $R$ plotted as a function of normalized time $T$.
Expanding Eq. \eqref{eq:RT} for $T\ll 1$ leads to 
\begin{equation}
R\simeq 1-\left(\frac T 3\right)^{1/2} \quad\text{for}\quad T\ll 1
\end{equation}
in accord with Lucas-Washburn theory, where the distance penetrated scales with the square root of time.
For late times the front $R$ approaches the center of the sphere ($R=0$) as
\begin{equation}
R\simeq \left(\frac{{1-T}}{{3}}\right)^{1/2}\quad\text{for}\quad T\lesssim 1.
\end{equation}


To facilitate a direct quantitative comparison with the experimental data (Fig. \ref{Fig:RawData}), we consider the mass of $m(t)$ of the seed 
\begin{equation}
m(t) = m_0 + \frac{4\pi}{3} \left(a^3- r_f^3 \right) \rho_0 \varepsilon \label{eq:m}
\end{equation}
where $m_0$ is the initial (dry) mass, $\rho_0$ is the liquid density, and $\varepsilon$ is the pore volume fraction. The maximum mass is
\begin{equation}
\label{mmax}
m_{max} = m_0 + \frac{4\pi}{3} a^3 \rho_0\varepsilon,
\end{equation}
which is attained at $t=t_{max}$. By normalizing Eq. \eqref{eq:m} by this value we get the dimensionless added mass
\begin{equation}
M = \frac{m(t) - m_0}{m_{max}-m_0}= 1- \left(\frac{r_f}{a}\right)^3=1-R(T)^3. \label{eq:M}
\end{equation}
Figure \ref{Fig:theory_radius_mass} shows the normalized added mass $M$ plotted as a function of normalized time $T$. Expanding Eq. \eqref{eq:M} around $T=0$ we find that 
\begin{eqnarray}
M(T) &\simeq \sqrt{3T} - \frac 23 T,
\end{eqnarray}
which we note is a remarkably good approximation that only differs from the full solution by $6\%$ at $T=1$.


\begin{figure}
\centering
\includegraphics[width=8cm]{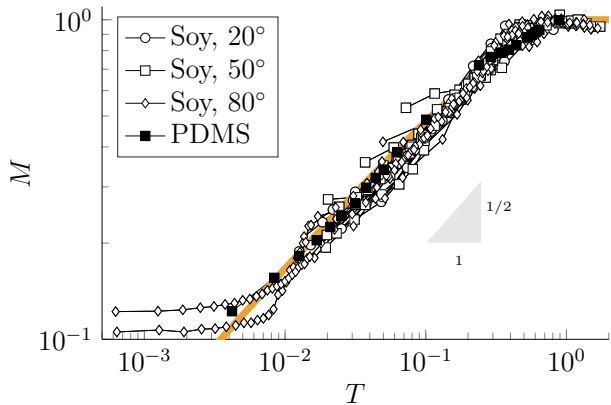}
\caption{Quantitative comparison between theory and experiment. The normalized mass $M=(m(t)-m_0)/(m_{\text{max}}-m_0)$ plotted as a function of normalized time $T=T/T_{\text{max}}$ for real and artificial seeds (points connected by lines).  The thick solid line shows the theoretical prediction (Eq. \eqref{eq:M}).  We observe reasonable agreement between theory and experiment, except for early times $T\sim 10^{-3}$ where the imbibition process is dominated by the hydration of the seed coat. The data in Fig. \ref{Fig:RawData} were fitted to Eq. \eqref{eq:M} using least squares to determine the best estimates of the parameters $t_{\text{max}},m_{\text{max}},$ and $m_0$.\label{Fig:ScaledData}} 
\end{figure}

The dynamical change in seed mass with time due to liquid imbibition predicted by Eqs. \eqref{eq:M} and \eqref{eq:RT} is compared to the experimentally obtained values in Fig. \ref{Fig:ScaledData}
 When normalized according to Eq. \eqref{eq:m}, the data obtained on both natural and artificial seeds collapse to a single curve. We observe reasonable agreement between theory (solid line) and experiments, except for early times $T\sim 10^{-3}$ where the imbibition process is dominated by the hydration of the seed coat. Moreover, the data agrees qualitatively with the tomographic images shown in Fig. \ref{Fig:Intro}(d): The iodine solution provided sufficient contrast to qualitatively follow the propagation of the liquid front inside the seed. After a delay during which the seed coat was penetrated, liquid quickly entered the porous material. The rate of water uptake gradually reduced, in accord with Fig. \ref{Fig:RawData}(a). The approximate rotational symmetry of the scans indicate that the imbibition process in soy seeds is homogenous, suggestive of a single pore scale porous material. 

A critical parameter of this study is the time to complete imbibition $t_{\text{max}}$. 
To estimates values of $t_{\text{max}}$ relevant to our soy seed experiments, we use the following parameters:
$ a=5\,\text{mm}$, $r_p = 3\,\text{nm}$ \cite{carpita1979determination}, $\gamma = 72\,\text{mN/m}$ \cite{vargaftik1983international}, and $ \theta = 83^\circ$ \cite{ray1958wetting}.
Assuming further that the viscosity of the starch-gel is $\mu = 10\,\text{mPa s}$  \cite{lai1991physicochemical} and that the porous material consists of ideal cylindrical pores (corresponding to $k=r_p^2/8$) leads to $t_{\text{max}}\simeq 6\times 10^3\,\text s $, in rough accord with the observed value of $t_{\text{max}}\sim   10^4 \,\text s$. For the PDMS spheres and diisopropylamine, we use parameters $r_p=0.1\,\text{nm}$, $\theta = 150\degree$, $\gamma = 20\,\text{mN/m}$, leading to $t_{\text{max}}=6.9\times 10^4\,\text s $, in reasonable agreement with observations ($t_{\text{max}}\sim 10^5$ s).

\section{Discussion and conclusion}
A reasonably clear picture of the factors that influence liquid uptake in real and artificial  seeds has emerged. In both cases, liquid permeates the substrate at a rate which decreases gradually over time (Fig. \ref{Fig:RawData}). Tomographic imaging of soy seeds confirmed this by observation of the permeating liquid (Fig. \ref{Fig:Intro} (d)). To rationalize the experimental data, we developed a model based on capillary forces pulling the liquid into the dry porous material. The theory predicted the temporal evolution of the radius of the wet front $R$ (Eq. \eqref{eq:RT}) and the seed mass $M$ (Eq. \eqref{eq:M}). Our model also showed that the time to fill the seed scales with the system parameters as $t_{\text{max}}\propto \mu a^2 /(r_p \gamma )$, i.e. an increase with the liquid viscosity $\mu$ and seed radius $a$, and a decrease with the pore size $r_p$ and surface tension $\gamma$.

The data obtained on both natural and artificial seeds collapse onto a single curve that agrees well with our model (Fig. \ref{Fig:ScaledData}). The observed agreement between theory and experiments conducted on both natural and artificial systems suggest that imbibition in soy seeds is driven -- at least initially -- primarily by capillary forces. At later times, and for other seed morphologies, however, effects such as growth of the developing embryo or tissue swelling may influence the process \cite{ha2018poro, kvick2017imbibition}. 


\begin{thebibliography}{29}%
\makeatletter
\providecommand \@ifxundefined [1]{%
 \@ifx{#1\undefined}
}%
\providecommand \@ifnum [1]{%
 \ifnum #1\expandafter \@firstoftwo
 \else \expandafter \@secondoftwo
 \fi
}%
\providecommand \@ifx [1]{%
 \ifx #1\expandafter \@firstoftwo
 \else \expandafter \@secondoftwo
 \fi
}%
\providecommand \natexlab [1]{#1}%
\providecommand \enquote  [1]{``#1''}%
\providecommand \bibnamefont  [1]{#1}%
\providecommand \bibfnamefont [1]{#1}%
\providecommand \citenamefont [1]{#1}%
\providecommand \href@noop [0]{\@secondoftwo}%
\providecommand \href [0]{\begingroup \@sanitize@url \@href}%
\providecommand \@href[1]{\@@startlink{#1}\@@href}%
\providecommand \@@href[1]{\endgroup#1\@@endlink}%
\providecommand \@sanitize@url [0]{\catcode `\\12\catcode `\$12\catcode
  `\&12\catcode `\#12\catcode `\^12\catcode `\_12\catcode `\%12\relax}%
\providecommand \@@startlink[1]{}%
\providecommand \@@endlink[0]{}%
\providecommand \url  [0]{\begingroup\@sanitize@url \@url }%
\providecommand \@url [1]{\endgroup\@href {#1}{\urlprefix }}%
\providecommand \urlprefix  [0]{URL }%
\providecommand \Eprint [0]{\href }%
\providecommand \doibase [0]{http://dx.doi.org/}%
\providecommand \selectlanguage [0]{\@gobble}%
\providecommand \bibinfo  [0]{\@secondoftwo}%
\providecommand \bibfield  [0]{\@secondoftwo}%
\providecommand \translation [1]{[#1]}%
\providecommand \BibitemOpen [0]{}%
\providecommand \bibitemStop [0]{}%
\providecommand \bibitemNoStop [0]{.\EOS\space}%
\providecommand \EOS [0]{\spacefactor3000\relax}%
\providecommand \BibitemShut  [1]{\csname bibitem#1\endcsname}%
\let\auto@bib@innerbib\@empty
\bibitem [{\citenamefont {Xiao}\ \emph {et~al.}(2012)\citenamefont {Xiao},
  \citenamefont {Stone},\ and\ \citenamefont {Attinger}}]{Xiao2012}%
  \BibitemOpen
  \bibfield  {author} {\bibinfo {author} {\bibfnamefont {J.}~\bibnamefont
  {Xiao}}, \bibinfo {author} {\bibfnamefont {H.~A.}\ \bibnamefont {Stone}}, \
  and\ \bibinfo {author} {\bibfnamefont {D.}~\bibnamefont {Attinger}},\
  }\href@noop {} {\bibfield  {journal} {\bibinfo  {journal} {Langmuir}\
  }\textbf {\bibinfo {volume} {28}},\ \bibinfo {pages} {4208} (\bibinfo {year}
  {2012})}\BibitemShut {NoStop}%
\bibitem [{\citenamefont {Bewley}(1997)}]{bewley1997seed}%
  \BibitemOpen
  \bibfield  {author} {\bibinfo {author} {\bibfnamefont {J.~D.}\ \bibnamefont
  {Bewley}},\ }\href@noop {} {\bibfield  {journal} {\bibinfo  {journal} {The
  plant cell}\ }\textbf {\bibinfo {volume} {9}},\ \bibinfo {pages} {1055}
  (\bibinfo {year} {1997})}\BibitemShut {NoStop}%
\bibitem [{\citenamefont {Linkies}\ \emph {et~al.}(2010)\citenamefont
  {Linkies}, \citenamefont {Graeber}, \citenamefont {Knight},\ and\
  \citenamefont {Leubner-Metzger}}]{linkies2010evolution}%
  \BibitemOpen
  \bibfield  {author} {\bibinfo {author} {\bibfnamefont {A.}~\bibnamefont
  {Linkies}}, \bibinfo {author} {\bibfnamefont {K.}~\bibnamefont {Graeber}},
  \bibinfo {author} {\bibfnamefont {C.}~\bibnamefont {Knight}}, \ and\ \bibinfo
  {author} {\bibfnamefont {G.}~\bibnamefont {Leubner-Metzger}},\ }\href@noop {}
  {\bibfield  {journal} {\bibinfo  {journal} {New Phytologist}\ }\textbf
  {\bibinfo {volume} {186}},\ \bibinfo {pages} {817} (\bibinfo {year}
  {2010})}\BibitemShut {NoStop}%
\bibitem [{\citenamefont {Bewley}\ \emph {et~al.}(2012)\citenamefont {Bewley},
  \citenamefont {Bradford}, \citenamefont {Hilhorst},\ and\ \citenamefont
  {hiroyuki nonogaki}}]{1461446929}%
  \BibitemOpen
  \bibfield  {author} {\bibinfo {author} {\bibfnamefont {J.~D.}\ \bibnamefont
  {Bewley}}, \bibinfo {author} {\bibfnamefont {K.}~\bibnamefont {Bradford}},
  \bibinfo {author} {\bibfnamefont {H.}~\bibnamefont {Hilhorst}}, \ and\
  \bibinfo {author} {\bibnamefont {hiroyuki nonogaki}},\ }\href@noop {} {\emph
  {\bibinfo {title} {Seeds: Physiology of Development, Germination and
  Dormancy, 3rd Edition}}}\ (\bibinfo  {publisher} {Springer},\ \bibinfo {year}
  {2012})\BibitemShut {NoStop}%
\bibitem [{\citenamefont {Swanson}\ \emph {et~al.}(1985)\citenamefont
  {Swanson}, \citenamefont {Hughes},\ and\ \citenamefont
  {Rasmussen}}]{swanson1985seed}%
  \BibitemOpen
  \bibfield  {author} {\bibinfo {author} {\bibfnamefont {B.~G.}\ \bibnamefont
  {Swanson}}, \bibinfo {author} {\bibfnamefont {J.~S.}\ \bibnamefont {Hughes}},
  \ and\ \bibinfo {author} {\bibfnamefont {H.~P.}\ \bibnamefont {Rasmussen}},\
  }\href@noop {} {\bibfield  {journal} {\bibinfo  {journal} {Food Structure}\
  }\textbf {\bibinfo {volume} {4}},\ \bibinfo {pages} {14} (\bibinfo {year}
  {1985})}\BibitemShut {NoStop}%
\bibitem [{\citenamefont {Nakayama}\ and\ \citenamefont
  {Komatsu}(2008)}]{nakayama2008water}%
  \BibitemOpen
  \bibfield  {author} {\bibinfo {author} {\bibfnamefont {N.}~\bibnamefont
  {Nakayama}}\ and\ \bibinfo {author} {\bibfnamefont {S.}~\bibnamefont
  {Komatsu}},\ }\href@noop {} {\bibfield  {journal} {\bibinfo  {journal} {Plant
  Production Science}\ }\textbf {\bibinfo {volume} {11}},\ \bibinfo {pages}
  {415} (\bibinfo {year} {2008})}\BibitemShut {NoStop}%
\bibitem [{\citenamefont {Sims}(1959)}]{sims1959germination}%
  \BibitemOpen
  \bibfield  {author} {\bibinfo {author} {\bibfnamefont {R.}~\bibnamefont
  {Sims}},\ }\href@noop {} {\bibfield  {journal} {\bibinfo  {journal} {Journal
  of the Institute of Brewing}\ }\textbf {\bibinfo {volume} {65}},\ \bibinfo
  {pages} {46} (\bibinfo {year} {1959})}\BibitemShut {NoStop}%
\bibitem [{\citenamefont {Ashraf}\ and\ \citenamefont
  {Abu-Shakra}(1978)}]{ashraf1978wheat}%
  \BibitemOpen
  \bibfield  {author} {\bibinfo {author} {\bibfnamefont {C.}~\bibnamefont
  {Ashraf}}\ and\ \bibinfo {author} {\bibfnamefont {S.}~\bibnamefont
  {Abu-Shakra}},\ }\href@noop {} {\bibfield  {journal} {\bibinfo  {journal}
  {Agronomy Journal}\ }\textbf {\bibinfo {volume} {70}},\ \bibinfo {pages}
  {135} (\bibinfo {year} {1978})}\BibitemShut {NoStop}%
\bibitem [{\citenamefont {Montanuci}\ \emph {et~al.}(2013)\citenamefont
  {Montanuci}, \citenamefont {Jorge},\ and\ \citenamefont
  {Jorge}}]{montanuci2013kinetic}%
  \BibitemOpen
  \bibfield  {author} {\bibinfo {author} {\bibfnamefont {F.~D.}\ \bibnamefont
  {Montanuci}}, \bibinfo {author} {\bibfnamefont {L.~M. d.~M.}\ \bibnamefont
  {Jorge}}, \ and\ \bibinfo {author} {\bibfnamefont {R.~M.~M.}\ \bibnamefont
  {Jorge}},\ }\href@noop {} {\bibfield  {journal} {\bibinfo  {journal} {Food
  Science and Technology (Campinas)}\ }\textbf {\bibinfo {volume} {33}},\
  \bibinfo {pages} {690} (\bibinfo {year} {2013})}\BibitemShut {NoStop}%
\bibitem [{\citenamefont {Rathjen}\ \emph {et~al.}(2009)\citenamefont
  {Rathjen}, \citenamefont {Strounina},\ and\ \citenamefont
  {Mares}}]{rathjen2009water}%
  \BibitemOpen
  \bibfield  {author} {\bibinfo {author} {\bibfnamefont {J.~R.}\ \bibnamefont
  {Rathjen}}, \bibinfo {author} {\bibfnamefont {E.~V.}\ \bibnamefont
  {Strounina}}, \ and\ \bibinfo {author} {\bibfnamefont {D.~J.}\ \bibnamefont
  {Mares}},\ }\href@noop {} {\bibfield  {journal} {\bibinfo  {journal} {Journal
  of experimental botany}\ }\textbf {\bibinfo {volume} {60}},\ \bibinfo {pages}
  {1619} (\bibinfo {year} {2009})}\BibitemShut {NoStop}%
\bibitem [{\citenamefont {Ha}\ \emph {et~al.}(2018)\citenamefont {Ha},
  \citenamefont {Kim}, \citenamefont {Jung}, \citenamefont {Yun}, \citenamefont
  {Kim},\ and\ \citenamefont {Kim}}]{ha2018poro}%
  \BibitemOpen
  \bibfield  {author} {\bibinfo {author} {\bibfnamefont {J.}~\bibnamefont
  {Ha}}, \bibinfo {author} {\bibfnamefont {J.}~\bibnamefont {Kim}}, \bibinfo
  {author} {\bibfnamefont {Y.}~\bibnamefont {Jung}}, \bibinfo {author}
  {\bibfnamefont {G.}~\bibnamefont {Yun}}, \bibinfo {author} {\bibfnamefont
  {D.-N.}\ \bibnamefont {Kim}}, \ and\ \bibinfo {author} {\bibfnamefont
  {H.-Y.}\ \bibnamefont {Kim}},\ }\href@noop {} {\bibfield  {journal} {\bibinfo
   {journal} {Science advances}\ }\textbf {\bibinfo {volume} {4}},\ \bibinfo
  {pages} {eaao7051} (\bibinfo {year} {2018})}\BibitemShut {NoStop}%
\bibitem [{\citenamefont {Gibson}(2012)}]{gibson2012hierarchical}%
  \BibitemOpen
  \bibfield  {author} {\bibinfo {author} {\bibfnamefont {L.~J.}\ \bibnamefont
  {Gibson}},\ }\href@noop {} {\bibfield  {journal} {\bibinfo  {journal}
  {Journal of the Royal Society Interface}\ ,\ \bibinfo {pages} {rsif20120341}}
  (\bibinfo {year} {2012})}\BibitemShut {NoStop}%
\bibitem [{\citenamefont {Kim}\ \emph {et~al.}(2017)\citenamefont {Kim},
  \citenamefont {Ha},\ and\ \citenamefont {Kim}}]{kim2017capillary}%
  \BibitemOpen
  \bibfield  {author} {\bibinfo {author} {\bibfnamefont {J.}~\bibnamefont
  {Kim}}, \bibinfo {author} {\bibfnamefont {J.}~\bibnamefont {Ha}}, \ and\
  \bibinfo {author} {\bibfnamefont {H.-Y.}\ \bibnamefont {Kim}},\ }\href@noop
  {} {\bibfield  {journal} {\bibinfo  {journal} {Journal of Fluid Mechanics}\
  }\textbf {\bibinfo {volume} {818}} (\bibinfo {year} {2017})}\BibitemShut
  {NoStop}%
\bibitem [{\citenamefont {Debacker}\ \emph {et~al.}(2014)\citenamefont
  {Debacker}, \citenamefont {Makarchuk}, \citenamefont {Lootens},\ and\
  \citenamefont {H{\'e}braud}}]{debacker2014imbibition}%
  \BibitemOpen
  \bibfield  {author} {\bibinfo {author} {\bibfnamefont {A.}~\bibnamefont
  {Debacker}}, \bibinfo {author} {\bibfnamefont {S.}~\bibnamefont {Makarchuk}},
  \bibinfo {author} {\bibfnamefont {D.}~\bibnamefont {Lootens}}, \ and\
  \bibinfo {author} {\bibfnamefont {P.}~\bibnamefont {H{\'e}braud}},\
  }\href@noop {} {\bibfield  {journal} {\bibinfo  {journal} {Physical review
  letters}\ }\textbf {\bibinfo {volume} {113}},\ \bibinfo {pages} {028301}
  (\bibinfo {year} {2014})}\BibitemShut {NoStop}%
\bibitem [{\citenamefont {Hong}\ \emph {et~al.}(2009)\citenamefont {Hong},
  \citenamefont {Hong}, \citenamefont {Lee}, \citenamefont {Cho}, \citenamefont
  {Lee}, \citenamefont {Cheong},\ and\ \citenamefont {Lee}}]{hong2009mr}%
  \BibitemOpen
  \bibfield  {author} {\bibinfo {author} {\bibfnamefont {Y.-S.}\ \bibnamefont
  {Hong}}, \bibinfo {author} {\bibfnamefont {K.~S.}\ \bibnamefont {Hong}},
  \bibinfo {author} {\bibfnamefont {E.-S.}\ \bibnamefont {Lee}}, \bibinfo
  {author} {\bibfnamefont {J.-H.}\ \bibnamefont {Cho}}, \bibinfo {author}
  {\bibfnamefont {C.}~\bibnamefont {Lee}}, \bibinfo {author} {\bibfnamefont
  {C.}~\bibnamefont {Cheong}}, \ and\ \bibinfo {author} {\bibfnamefont {C.-H.}\
  \bibnamefont {Lee}},\ }\href@noop {} {\bibfield  {journal} {\bibinfo
  {journal} {Food research international}\ }\textbf {\bibinfo {volume} {42}},\
  \bibinfo {pages} {237} (\bibinfo {year} {2009})}\BibitemShut {NoStop}%
\bibitem [{\citenamefont {Gruwel}\ \emph {et~al.}(2001)\citenamefont {Gruwel},
  \citenamefont {Chatson}, \citenamefont {Yin},\ and\ \citenamefont
  {Abrams}}]{gruwel2001magnetic}%
  \BibitemOpen
  \bibfield  {author} {\bibinfo {author} {\bibfnamefont {M.~L.}\ \bibnamefont
  {Gruwel}}, \bibinfo {author} {\bibfnamefont {B.}~\bibnamefont {Chatson}},
  \bibinfo {author} {\bibfnamefont {X.~S.}\ \bibnamefont {Yin}}, \ and\
  \bibinfo {author} {\bibfnamefont {S.}~\bibnamefont {Abrams}},\ }\href@noop {}
  {\bibfield  {journal} {\bibinfo  {journal} {International journal of food
  science \& technology}\ }\textbf {\bibinfo {volume} {36}},\ \bibinfo {pages}
  {161} (\bibinfo {year} {2001})}\BibitemShut {NoStop}%
\bibitem [{\citenamefont {Gart}\ \emph {et~al.}(2015)\citenamefont {Gart},
  \citenamefont {Mates}, \citenamefont {Megaridis},\ and\ \citenamefont
  {Jung}}]{gart2015}%
  \BibitemOpen
  \bibfield  {author} {\bibinfo {author} {\bibfnamefont {S.}~\bibnamefont
  {Gart}}, \bibinfo {author} {\bibfnamefont {J.~E.}\ \bibnamefont {Mates}},
  \bibinfo {author} {\bibfnamefont {C.~M.}\ \bibnamefont {Megaridis}}, \ and\
  \bibinfo {author} {\bibfnamefont {S.}~\bibnamefont {Jung}},\ }\href@noop {}
  {\bibfield  {journal} {\bibinfo  {journal} {Physicl Review Applied}\ }\textbf
  {\bibinfo {volume} {3}},\ \bibinfo {pages} {044019} (\bibinfo {year}
  {2015})}\BibitemShut {NoStop}%
\bibitem [{\citenamefont {Ludwig}\ \emph {et~al.}(2009)\citenamefont {Ludwig},
  \citenamefont {King}, \citenamefont {Reischig}, \citenamefont {Herbig},
  \citenamefont {Lauridsen}, \citenamefont {Schmidt}, \citenamefont {Proudhon},
  \citenamefont {Forest}, \citenamefont {Cloetens}, \citenamefont {Du~Roscoat}
  \emph {et~al.}}]{ludwig2009new}%
  \BibitemOpen
  \bibfield  {author} {\bibinfo {author} {\bibfnamefont {W.}~\bibnamefont
  {Ludwig}}, \bibinfo {author} {\bibfnamefont {A.}~\bibnamefont {King}},
  \bibinfo {author} {\bibfnamefont {P.}~\bibnamefont {Reischig}}, \bibinfo
  {author} {\bibfnamefont {M.}~\bibnamefont {Herbig}}, \bibinfo {author}
  {\bibfnamefont {E.~M.}\ \bibnamefont {Lauridsen}}, \bibinfo {author}
  {\bibfnamefont {S.}~\bibnamefont {Schmidt}}, \bibinfo {author} {\bibfnamefont
  {H.}~\bibnamefont {Proudhon}}, \bibinfo {author} {\bibfnamefont
  {S.}~\bibnamefont {Forest}}, \bibinfo {author} {\bibfnamefont
  {P.}~\bibnamefont {Cloetens}}, \bibinfo {author} {\bibfnamefont {S.~R.}\
  \bibnamefont {Du~Roscoat}},  \emph {et~al.},\ }\href@noop {} {\bibfield
  {journal} {\bibinfo  {journal} {Materials Science and Engineering: A}\
  }\textbf {\bibinfo {volume} {524}},\ \bibinfo {pages} {69} (\bibinfo {year}
  {2009})}\BibitemShut {NoStop}%
\bibitem [{\citenamefont {Feldkamp}\ \emph {et~al.}(1984)\citenamefont
  {Feldkamp}, \citenamefont {Davis},\ and\ \citenamefont
  {Kress}}]{feldkamp1984practical}%
  \BibitemOpen
  \bibfield  {author} {\bibinfo {author} {\bibfnamefont {L.}~\bibnamefont
  {Feldkamp}}, \bibinfo {author} {\bibfnamefont {L.}~\bibnamefont {Davis}}, \
  and\ \bibinfo {author} {\bibfnamefont {J.}~\bibnamefont {Kress}},\
  }\href@noop {} {\bibfield  {journal} {\bibinfo  {journal} {JOSA A}\ }\textbf
  {\bibinfo {volume} {1}},\ \bibinfo {pages} {612} (\bibinfo {year}
  {1984})}\BibitemShut {NoStop}%
\bibitem [{\citenamefont {Lucas}(1918)}]{lucas1918rate}%
  \BibitemOpen
  \bibfield  {author} {\bibinfo {author} {\bibfnamefont {R.}~\bibnamefont
  {Lucas}},\ }\href@noop {} {\bibfield  {journal} {\bibinfo  {journal} {Kolloid
  Z}\ }\textbf {\bibinfo {volume} {23}},\ \bibinfo {pages} {15} (\bibinfo
  {year} {1918})}\BibitemShut {NoStop}%
\bibitem [{\citenamefont {Washburn}(1921)}]{washburn1921dynamics}%
  \BibitemOpen
  \bibfield  {author} {\bibinfo {author} {\bibfnamefont {E.~W.}\ \bibnamefont
  {Washburn}},\ }\href@noop {} {\bibfield  {journal} {\bibinfo  {journal}
  {Physical review}\ }\textbf {\bibinfo {volume} {17}},\ \bibinfo {pages} {273}
  (\bibinfo {year} {1921})}\BibitemShut {NoStop}%
\bibitem [{\citenamefont {de~Gennes}\ \emph {et~al.}(2010)\citenamefont
  {de~Gennes}, \citenamefont {Brochard-Wyart},\ and\ \citenamefont
  {Quere}}]{Capillarity}%
  \BibitemOpen
  \bibfield  {author} {\bibinfo {author} {\bibfnamefont {P.-G.}\ \bibnamefont
  {de~Gennes}}, \bibinfo {author} {\bibfnamefont {F.}~\bibnamefont
  {Brochard-Wyart}}, \ and\ \bibinfo {author} {\bibfnamefont {D.}~\bibnamefont
  {Quere}},\ }\href@noop {} {\emph {\bibinfo {title} {Capillarity and Wetting
  Phenomena: Drops, Bubbles, Pearls, Waves}}}\ (\bibinfo  {publisher}
  {Springer},\ \bibinfo {year} {2010})\BibitemShut {NoStop}%
\bibitem [{\citenamefont {Bohin}\ \emph {et~al.}(1994)\citenamefont {Bohin},
  \citenamefont {Manas-Zloczower},\ and\ \citenamefont
  {Feke}}]{bohin1994penetration}%
  \BibitemOpen
  \bibfield  {author} {\bibinfo {author} {\bibfnamefont {F.}~\bibnamefont
  {Bohin}}, \bibinfo {author} {\bibfnamefont {I.}~\bibnamefont
  {Manas-Zloczower}}, \ and\ \bibinfo {author} {\bibfnamefont {D.}~\bibnamefont
  {Feke}},\ }\href@noop {} {\bibfield  {journal} {\bibinfo  {journal} {Rubber
  chemistry and technology}\ }\textbf {\bibinfo {volume} {67}},\ \bibinfo
  {pages} {602} (\bibinfo {year} {1994})}\BibitemShut {NoStop}%
\bibitem [{\citenamefont {Bohin}\ \emph {et~al.}(1995)\citenamefont {Bohin},
  \citenamefont {Feke},\ and\ \citenamefont
  {Manas-Zloczower}}]{bohin1995determination}%
  \BibitemOpen
  \bibfield  {author} {\bibinfo {author} {\bibfnamefont {F.}~\bibnamefont
  {Bohin}}, \bibinfo {author} {\bibfnamefont {D.}~\bibnamefont {Feke}}, \ and\
  \bibinfo {author} {\bibfnamefont {I.}~\bibnamefont {Manas-Zloczower}},\
  }\href@noop {} {\bibfield  {journal} {\bibinfo  {journal} {Powder
  technology}\ }\textbf {\bibinfo {volume} {83}},\ \bibinfo {pages} {159}
  (\bibinfo {year} {1995})}\BibitemShut {NoStop}%
\bibitem [{\citenamefont {Carpita}\ \emph {et~al.}(1979)\citenamefont
  {Carpita}, \citenamefont {Sabularse}, \citenamefont {Montezinos},\ and\
  \citenamefont {Delmer}}]{carpita1979determination}%
  \BibitemOpen
  \bibfield  {author} {\bibinfo {author} {\bibfnamefont {N.}~\bibnamefont
  {Carpita}}, \bibinfo {author} {\bibfnamefont {D.}~\bibnamefont {Sabularse}},
  \bibinfo {author} {\bibfnamefont {D.}~\bibnamefont {Montezinos}}, \ and\
  \bibinfo {author} {\bibfnamefont {D.~P.}\ \bibnamefont {Delmer}},\
  }\href@noop {} {\bibfield  {journal} {\bibinfo  {journal} {Science}\ }\textbf
  {\bibinfo {volume} {205}},\ \bibinfo {pages} {1144} (\bibinfo {year}
  {1979})}\BibitemShut {NoStop}%
\bibitem [{\citenamefont {Vargaftik}\ \emph {et~al.}(1983)\citenamefont
  {Vargaftik}, \citenamefont {Volkov},\ and\ \citenamefont
  {Voljak}}]{vargaftik1983international}%
  \BibitemOpen
  \bibfield  {author} {\bibinfo {author} {\bibfnamefont {N.}~\bibnamefont
  {Vargaftik}}, \bibinfo {author} {\bibfnamefont {B.}~\bibnamefont {Volkov}}, \
  and\ \bibinfo {author} {\bibfnamefont {L.}~\bibnamefont {Voljak}},\
  }\href@noop {} {\bibfield  {journal} {\bibinfo  {journal} {Journal of
  Physical and Chemical Reference Data}\ }\textbf {\bibinfo {volume} {12}},\
  \bibinfo {pages} {817} (\bibinfo {year} {1983})}\BibitemShut {NoStop}%
\bibitem [{\citenamefont {Ray}\ \emph {et~al.}(1958)\citenamefont {Ray},
  \citenamefont {Anderson},\ and\ \citenamefont {Scholz}}]{ray1958wetting}%
  \BibitemOpen
  \bibfield  {author} {\bibinfo {author} {\bibfnamefont {B.~R.}\ \bibnamefont
  {Ray}}, \bibinfo {author} {\bibfnamefont {J.}~\bibnamefont {Anderson}}, \
  and\ \bibinfo {author} {\bibfnamefont {J.}~\bibnamefont {Scholz}},\
  }\href@noop {} {\bibfield  {journal} {\bibinfo  {journal} {The Journal of
  Physical Chemistry}\ }\textbf {\bibinfo {volume} {62}},\ \bibinfo {pages}
  {1220} (\bibinfo {year} {1958})}\BibitemShut {NoStop}%
\bibitem [{\citenamefont {Lai}\ and\ \citenamefont
  {Kokini}(1991)}]{lai1991physicochemical}%
  \BibitemOpen
  \bibfield  {author} {\bibinfo {author} {\bibfnamefont {L.}~\bibnamefont
  {Lai}}\ and\ \bibinfo {author} {\bibfnamefont {J.}~\bibnamefont {Kokini}},\
  }\href@noop {} {\bibfield  {journal} {\bibinfo  {journal} {Biotechnology
  progress}\ }\textbf {\bibinfo {volume} {7}},\ \bibinfo {pages} {251}
  (\bibinfo {year} {1991})}\BibitemShut {NoStop}%
\bibitem [{\citenamefont {Kvick}\ \emph {et~al.}(2017)\citenamefont {Kvick},
  \citenamefont {Martinez}, \citenamefont {Hewitt},\ and\ \citenamefont
  {Balmforth}}]{kvick2017imbibition}%
  \BibitemOpen
  \bibfield  {author} {\bibinfo {author} {\bibfnamefont {M.}~\bibnamefont
  {Kvick}}, \bibinfo {author} {\bibfnamefont {D.~M.}\ \bibnamefont {Martinez}},
  \bibinfo {author} {\bibfnamefont {D.~R.}\ \bibnamefont {Hewitt}}, \ and\
  \bibinfo {author} {\bibfnamefont {N.~J.}\ \bibnamefont {Balmforth}},\
  }\href@noop {} {\bibfield  {journal} {\bibinfo  {journal} {Physical Review
  Fluids}\ }\textbf {\bibinfo {volume} {2}},\ \bibinfo {pages} {074001}
  (\bibinfo {year} {2017})}\BibitemShut {NoStop}%
\end{thebibliography}
\end{document}